\documentclass[showpacs,preprintnumbers,amsmath,amssymb]{revtex4}

\usepackage{graphicx}
\usepackage{dcolumn}
\usepackage{bm}

\usepackage{latexsym}
\usepackage{amssymb}
\usepackage{amsfonts}
\usepackage{amsmath}
\usepackage{theorem}

\newtheorem{proposition}{Proposition}
\newtheorem{lemma}{Lemma}
\newtheorem{remark}{Remark}
\newtheorem{theorem}{Theorem}
 \newtheorem{corollary}{Corollary}

\begin{document}

\title{\bf About the ergodic regime in the analogical Hopfield neural networks.\\ Moments of the partition function}

\author{Adriano Barra$^{*\dag}$}
\author{Francesco Guerra$^{*\ddag}$}
\affiliation{$^*$ Dipartimento di Fisica, Sapienza Universit\`a di
Roma, Piazzale Aldo Moro 2, 00185 Roma, Italy \\
$^\dag$ Dipartimento di Matematica, Universit\`a di Bologna,
Piazza San Donato 5, 40126 Bologna, Italy \\
$\ddag$ INFN, Sezione Roma1, Piazzale Aldo Moro 2, 00185 Roma, Italy  
                           \email{Adriano.Barra@roma1.infn.it, \ Francesco.Guerra@roma1.infn.it}}
\def\be{\begin{equation}}
\def\ee{\end{equation}}
\def\bc{\begin{center}}
\def\ec{\end{center}}

\date{Received: date / Accepted: date}

\date{\today}

\begin{abstract}
In this paper we introduce and exploit the real replica approach for a minimal generalization of
the Hopfield model, by assuming the learned patterns  to be distributed accordingly to a standard unit
Gaussian. We consider the
high storage case, when the number of patterns is linearly diverging with the number of neurons. We study the infinite volume behavior of the normalized momenta of the partition function. We find a region in the parameter space where the free energy density in the infinite volume limit is self-averaging  around its annealed approximation, as well as the
entropy and the internal energy density. Moreover, we evaluate the corrections
to their extensive counterparts with respect to their annealed
expressions. The fluctuations of properly introduced overlaps,
which act as order parameters, are also discussed. 
\end{abstract}

\maketitle

\section{Introduction}\label{due}

In the last twenty years, from the early work by Hopfield
\cite{hopfield} and the, nowadays historical, theory of Amit,
Gutfreund and Sompolinsky (AGS) \cite{amit,ags1,ags2} to the
modern theory for learning \cite{hotel, peter}, the neural
networks, thought of as spin glasses with a Hebb-like ``synaptic
matrix'' \cite{hebb}, became more and more important in several
different contexts, as artificial intelligence \cite{AI}, cognitive
psychology \cite{alb3}, problem solving  \cite{alb2, alb4},
and so on. Despite their fundamental role, and due to their very
 difficult  mathematical control, very little is rigorously known
 about these networks. Along the years several contributions
 appeared  (\textit{e.g.} \cite{tirozzi1, bovier1, bovier2,bovier3, bovier4, tirozzi2, tirozzi3, talahopfield1, talahopfield2}), often
   motivated by an increasing
 understanding of spin-glasses  (\textit{e.g.} \cite{broken, guerra2, Gsum, pastur, Talabook}) and the analysis at low level of stored memories has been achieved.
\newline
However in the high level of stored memories, fundamental
enquiries are still rather incomplete. Furthermore, general problems as the
 existence of a well defined thermodynamic limit
  are unsolved, in contrast with the spin glass case \cite{limterm, limterm2}.
\newline
In this paper, we introduce some techniques (essentially in
the real replica framework) developed for the spin glass theory
(see \textit{e.g.} \cite{guerrasg, gt2, comets, barra2}),
in order to give a  description of the Hopfield model
in the high temperature region with high level of stored memories
(\textit{i.e.} patterns), by no use of replica trick \cite{MPV}. We take the
freedom of allowing the learned patterns to take all real values,
their probability distribution being a standard Gaussian
$\mathcal{N}[0,1]$, and we refer to this minimal generalization as
{\itshape analogical} Hopfield model, to stress that the memories
are no longer discrete as in standard literature.
\newline
Within this scenario, we exploit the moment method in order to prove bounds on the critical line for
the ergodic phase and give the explicit expression for all the
thermodynamical quantities in the infinite volume limit, in complete agreement with AGS theory.
Furthermore we show in a  simple way self-averaging of free and
internal energy and entropy per site and calculate their extensive
fluctuations around the annealed expressions, in analogy with what was found by Aizenman, Lebowitz and Ruelle \cite{ALR} for the Sherrington-Kirkpatrick model in the ergodic region. 
\newline
We investigate also about the overlap fluctuations, in the ergodic region. The paper is organized as follows. In section  
\ref{model} we define the analogical Hopfield model, and show that it is equivalent to a bipartite spin glass, where one party is described by Ising spins and the other by Gaussian spins. We introduce also the main thermodynamic quantities and their annealed approximation. In the next section \ref{overlap} we introduce overlaps for replicas of the Ising spins and the Gaussian spins, and show how they enter in the expression of thermodynamic quantities, as for example the internal energy. In section IV we state our main results about the validity of the annealed approximation, and establish the fluctuations of the extensive thermodynamic variables and the overlaps. In section V we study the momenta of the normalized partition function in the infinite volume limit, and prove our results about the annealed approximation. In section VI we prove the log-normality of the limiting distribution for the partition function, and prove the rest of our results. Finally, section VII is devoted to some conclusion and outlook for future developments.   

\section{Definition of the model}\label{model}

We introduce a large network of $N$ two-state neurons  $ (1,..,N)\ni i \to \sigma_i = \pm1$, which are thought of as quiescent
(sleeping) when their value is $-1$ or spiking (emitting a current
signal to other neurons) when their value is $+1$. They interact
throughout a synaptic matrix $J_{ij}$ defined according to the Hebb
rule for learning
\be J_{ij} = \sum_{\mu=1}^p \xi_i^{\mu}\xi_j^{\mu}. \ee
Each random variable $\xi^{\mu}=\{\xi_1^{\mu},..,\xi_N^{\mu}\}$
represents a learned pattern and tries to bring the overall
current in the network (or in some part) stable with respect to
itself (when this happens, we say we have a retrieval state, see
\textit{e.g.} \cite{amit}). The analysis of the network assumes that the
system has already stored $p$ patterns (no learning is
investigated here), and we will be interested in the case in which this
number increases linearly with respect to the system size
(high storage level), so that $p/N\to\alpha$ as $N\to\infty$, where $\alpha\ge0$ is a parameter of the theory denoting the storage level.
\newline
In standard literature these patterns are usually chosen at random
independently with values $\pm1$ taken with equal probability $1/2$. 
Here, we chose them as taking real values with a unit
Gaussian probability distribution, \textit{i.e.} \be d\mu(\xi_i^{\mu}) =
\frac{1}{\sqrt{2\pi}}e^{-(\xi_i^{\mu})^2/2}. \ee Of course,
avoiding pathological case, in the high storage level and in the
high temperature region, the results should show robustness  with
respect to the particular choice of the
probability distribution and we should recover the standard AGS
theory. The use of a Gaussian distribution has some technical advantages, as it allows to easily borrow powerful techniques from the spin glass case. The physical interpretation is very simple. While the patterns with $\pm1$ values describe ``images'' with white or black pixels, respectively, the patterns with continuous values describe ``gray'' pixels with a continuously variable luminosity from $-\infty$ (completely black) to $+\infty$ (completely white).   
\newline
The Hamiltonian of the model involves interactions between any couple of sites according to the definition \be H_N(\sigma;\xi) = - \frac{1}{N}\sum_{\mu=1}^p
\sum_{i<j}^N\xi_i^{\mu}\xi_j^{\mu}\sigma_i\sigma_j. \ee
By splitting the summations $\sum_{i<j}^N =
\frac{1}{2}\sum_{ij}^N - \frac12 \sum_i^N \delta_{ij}$, we can write  the partition function in the following form
\be\label{due}
Z_{N,p}(\beta;\xi) = \sum_{\sigma}
\exp{\Big(\frac{\beta}{2N}\sum_{\mu=1}^p\sum_{ij}^N
\xi_i^{\mu}\xi_j^{\mu}\sigma_i\sigma_j -
\frac{\beta}{2N}\sum_{\mu=1}^p\sum_{i}^N (\xi_i^{{\mu}})^2 \Big)}=
\tilde{Z}_{N,p}(\beta;\xi)e^{\frac{-\beta}{2N}\sum_{\mu=1}^p\sum_{i=1}^N
(\xi_i^{{\mu}})^2} \ee
where $\beta\ge0$ is the inverse temperature, and denotes here the level of noise in the network. We have defined
\be\label{Ztilde} \tilde{Z}_{N,p}(\beta;\xi)=
\sum_{\sigma}\exp(\frac{\beta}{2N}\sum_{\mu=1}^p\sum_{ij}^N
\xi_i^{\mu}\xi_j^{\mu}\sigma_i\sigma_j ). \ee Notice that the last
term at the r.h.s. of eq. (\ref{due}) does not depend on the
particular state of the network.
\newline
As a consequence, the control of the last term can be easily obtained. In fact, let us define the random variable $\hat{f}_N$ so that
\be\label{fcappuccio}
\hat{f}_N=\frac{1}{N}\sum_{\mu}^p\sum_{i}^N
(\xi_i^{\mu})^2.\ee
Then we have
\be\label{ZZtilde}
\ln Z_{N,p}(\beta;\xi)=    \ln \tilde{Z}_{N,p}(\beta;\xi) - \frac{\beta}{2}\hat{f}_N.
\ee
Since $\mathbb{E}\hat{f}_N=p$ we have immediately $\lim_{N\rightarrow\infty}(1/N)\mathbb{E}\hat{f}_N=\alpha$. On the other hand, $\hat{f}_N$ is a sum of independent random variables and therefore, by the strong law of large numbers, we have also $\lim_{N\rightarrow\infty}(1/N)\hat{f}_N=\alpha$, $\xi$-almost surely. 
\newline
Furthermore, by the central limit theorem,  we have, in distribution, $\lim_{N\rightarrow
\infty}(\hat{f}_N - \mathbb{E}\hat{f}_N) = \sqrt{2 \alpha}\ \chi$,
$\chi$ being a standard Gaussian $\mathcal{N}(0,1)$. In fact, \be
\mathbb{E}(\hat{f}^2_N)=\frac{1}{N^2}\sum_{\mu}^p \sum_i^N
\sum_{\nu}^p
\sum_j^N\mathbb{E}\Big((\xi_i^{\mu})^2(\xi_j^{\nu})^2\Big)= p^2 +
2 \frac{p}{N}, \ee so that
$\mathbb{E}(\hat{f}^2_N)-\mathbb{E}(\hat{f}_N)^2 = 2 p/N$, which
in the thermodynamic limit gives  the result.
\newline
Consequently we focus just on $\tilde{Z}(\beta;\xi)$. Let us apply
the Hubbard-Stratonovich lemma \cite{ellis} to linearize with
respect to the bilinear quenched memories carried by the
$\xi_i^{\mu}\xi_j^{\mu}$. If we define the ``Mattis
magnetization'' \cite{amit} $m_{\mu}$ as \be\label{emme} m_{\mu}=
\frac{1}{N} \sum_i^N \xi_i^{\mu}\sigma_i, \ee we can write
 \be\label{enne} \tilde{Z}_{N,p}(\beta;\xi) = \sum_{\sigma}\exp(\frac{\beta N}{2}\sum_{\mu=1}^p
 m_{\mu}^2) = \sum_{\sigma}\int  (\prod_{\mu=1}^p \frac{dz_{\mu}\exp(-z^2_{\mu}/2)}{\sqrt{2\pi}}) \exp(\sqrt{\beta N}\sum_{\mu=1}^p m^{\mu}z_{\mu}).
 \ee
Using eq. (\ref{enne}), the expression for the partition function
(\ref{due}) becomes \be\label{bipartito} Z_{N,p}(\beta;\xi) =
\sum_{\sigma}\int (\prod_{\mu=1}^p d\mu(z_{\mu}))
\exp(\sqrt{\frac{\beta}{N}}\sum_{\mu=1}^p \sum_{i=1}^N
\xi_i^{\mu}\sigma_i z_{\mu})\exp(-\frac{\beta}{2}\hat{f}_{N})
,\ee
with $d\mu(z_{\mu})$ the standard  Gaussian measure for all the
$z_{\mu}$.
\newline
For a generic function $F$ of the neurons, we define the
Boltzmann state $\omega_{\beta}(F)$ at a given level of noise
$\beta$ as the average \be \omega_{\beta}(F) = \omega(F)=
(Z_{N,p}(\beta;\xi))^{-1}\sum_{\sigma}F(\sigma)e^{-\beta
H_N(\sigma;\xi)} \ee and often we will drop the subscript $\beta$
for the sake of simplicity. Notice that the Boltzmann state does not involve the function $\hat{f}_N$, which factors out. The $s$-replicated Boltzmann state
is defined as the product state $\Omega = \omega^1\times \omega^2 \times ... \times
\omega^s$, in which all the single Boltzmann states are at the same noise level $\beta^{-1}$ and share an identical
sample of quenched memories $\xi$. For the sake of
clearness, given a function $F$ of the neurons of the $s$ replicas
and  using the symbol $a \in [1,..,s]$ to label
replicas, such an average can be written as
\be \Omega(F(\sigma^1,...,\sigma^s)) =
\frac{1}{Z_{N,p}^s}\sum_{\sigma^1}\sum_{\sigma^2}...\sum_{\sigma^s}
F(\sigma^1,...,\sigma^s)\exp(-\beta \sum_{a=1}^s
H_N(\sigma^{a},\xi)). \ee
The average over the quenched memories will be denoted by
$\mathbb{E}$ and for a generic function of these memories $F(\xi)$
 can be written as \be \mathbb{E}[F(\xi)] = \int
(\prod_{\mu=1}^p \prod_{i=1}^N \frac{d
\xi_i^{\mu}e^{-\frac{(\xi_i^{\mu})^2}{2}}}{\sqrt{2\pi}})F(\xi)=
\int F(\xi)d\mu(\xi), \ee of course $ \mathbb{E}[\xi_i^{\mu}]=0$
and $ \mathbb{E}[(\xi_i^{\mu})^2]=1$.
\newline
We use the symbol $\langle . \rangle$ to mean $\langle . \rangle =
\mathbb{E}\Omega(.)$.
\newline
Recall that in  the thermodynamic limit it is assumed
$$
\lim_{N \rightarrow \infty} \frac{p}{N}= \alpha,
$$
$\alpha$ being a given real number, which acts as free parameter of the theory.
\newline
The main quantity of interest is the intensive pressure
defined as \be A_{N,p}(\beta,\xi)= -\beta f_{N,p}(\beta,\xi) =
\frac{1}{N}\ln Z_{N,p}(\beta;\xi), \ee  while the quenched intensive pressure
is defined as 
\be A^*_{N,p}(\beta)= -\beta f^*_{N,p}(\beta) =
\frac{1}{N}\mathbb{E}\ln Z_{N,p}(\beta;\xi), \ee and the 
annealed intensive pressure is defined as  \be
\bar{A}_{N,p}(\beta) = -\beta \bar{f}_{N,p}(\beta) =
\frac{1}{N}\ln \mathbb{E} Z_{N,p}(\beta;\xi). \ee According to thermodynamics, here
$f_{N,p}(\beta,\xi)= u_{N,p}(\beta,\xi)-\beta^{-1}s_{N,p}(\beta, \xi)$
is the free energy density, $u_{N,p}(\beta,\xi)$ is the internal
energy density and $s_{N,p}(\beta,\xi)$ is the intensive entropy (the star and 
the bar  denote the quenched and the annealed evaluations as
well). Obviously, by Jensen inequality, we have $A^*_{N,p}(\beta)\le \bar{A}_{N,p}(\beta)$. 
%
\section{The role of the overlaps and the internal energy}\label{overlap}

According to the bipartite nature of the Hopfield model expressed by
 eq. (\ref{bipartito}),  we introduce two other order parameters
 beyond  the ``Mattis magnetization'' (eq. (\ref{emme})): the first is the overlap
between the replicated neurons, defined as \be q_{ab}= \frac1N
\sum_{i=1}^N \sigma_i^a \sigma_i^b \in [-1,+1], \ee and the second
the overlap between the replicated Gaussian  variables $z$,
defined as \be p_{ab} = \frac1p \sum_{\mu=1}^p z_{\mu}^{a}z_{\mu}^{b}
\in (-\infty, +\infty). \ee
These overlaps play a considerable role in the theory
as they can express thermodynamical quantities. As an example let
us work out the quenched internal energy of the model
$u^{*}_{N,p}(\beta)=N^{-1}\langle H_N(\sigma;\xi) \rangle$.
\begin{proposition}  For $\beta\ne1$,
the quenched internal energy $u^{*}_N(\alpha,\beta)$ of the analogical Hopfield model can be expressed as \be\label{acca}
u^{*}_{N,p}(\beta)=  \frac{\langle
H_N(\sigma;\xi) \rangle}{N} = - \frac{\beta}{2}(\frac{(p/N) -
\sum_{\mu}^p \langle m_{\mu}^1 q_{12} m_{\mu}^2
\rangle}{1-\beta}). \ee
\end{proposition}
\textbf{Proof}
\newline
The proof is based on direct calculations. We use Gaussian
integration and integration by parts over the Gaussian memories. Let us begin with
\begin{eqnarray}\label{euro0}
N^{-1}\mathbb{E}\Omega(H_N(\sigma;\xi)) &=&
-\frac{1}{2}\sum_{\mu}^p\mathbb{E}(\Omega(m_{\mu}^2))+\frac{1}{2N^2}\sum_{\mu}^p\sum_{i}^N
\mathbb{E}(\xi_i^{\mu 2}) \nonumber \\  &=&
-\frac{1}{2}\sum_{\mu}^p\mathbb{E}(\Omega(m_{\mu}^2)) +
\frac{p}{2N}. \\ \nonumber
\end{eqnarray}
Now we write explicitly a Mattis magnetization into the Boltzmann
average of (\ref{euro0}) so to use integration by parts over the
memories $\xi_i^{\mu}$ (i.e. $\mathbb{E}\xi
F(\xi)=\mathbb{E}\partial_{\xi} F(\xi)$).
\begin{eqnarray}\label{euro1}
-\frac{1}{2}\sum_{\mu}^p\mathbb{E}(\Omega(m_{\mu}^2)) +
\frac{p}{2N}
&=&-\frac{1}{2N}\sum_{\mu}^p\sum_i^N\mathbb{E}(\xi_i^{\mu}\Omega(\sigma_i
m_{\mu})) + \frac{p}{2N} \\ \nonumber &=&
-\frac{\beta}{2N}\sum_{\mu}^p\sum_i^N
\mathbb{E}\Big(\Omega((\sigma_i m_{\mu})^2) - \Omega^{2}(\sigma_i
m_{\mu}) + \frac{1}{\beta N} \Big) + \frac{p}{2N} \\
\nonumber &=& \beta \frac{\langle H(\sigma; \xi) \rangle}{N} -
\frac{\beta p}{2N}+ \frac{\beta}{2}\sum_{\mu}\langle
m_{\mu}^1 q_{12} m_{\mu}^2 \rangle \ \Box
\end{eqnarray}
In particular for $\beta=1$ we get exactly for any $N$ the very remarkable expression \be \sum_{\mu=1}^p \langle
m_{\mu}^1 q_{12} m_{\mu}^2 \rangle = \frac{p}{N}, \ee as it can be
understood by looking at the last line of (\ref{euro1}) and choosing
$\beta=1$.
\newline
\newline
At the end of this section we write down a short formulary in
which we consider the streaming of both the partition function and
the Boltzmann state with respect to the level of noise $\beta$ and
to a generic stored pattern $\xi_i^{\mu}$, as these calculations
will be useful several times along the paper.
\begin{eqnarray}\label{323}
\partial_{\beta}Z_{N,p}(\beta;\xi)
%
%
&=& \sum_{\sigma} \int (\prod_{\mu=1}^p d\mu(z^{\mu})) \sum_{i=1}^N
\sum_{\mu=1}^p \Big(
\frac{\xi_i^{\mu}\sigma_iz^{\mu}}{2\sqrt{\beta
N}}-\frac{(\xi_i^\mu)^2}{2N} \Big)\times \nonumber \\
&\times&\exp\Big[ \sum_{i=1}^N\sum_{\mu=1}^p\Big(
\sqrt{\frac{\beta}{N}}\xi_i^{\mu}\sigma_iz^{\mu}-\beta\frac{(\xi_i^{\mu})^2}{2N}
\Big) \Big]= \nonumber \\ &=&\sum_{i=1}^N\sum_{\mu=1}^p
(\frac{\omega(\xi_i^{\mu}\sigma_i z^{\mu})}{2 \sqrt{\beta N}} -
\frac{(\xi_i^{\mu})^2}{2N})Z_{N,p}(\beta;\xi) \\ \label{324}
\partial_{\xi_i^{\mu}} Z^{\lambda}_{N,p}(\beta;\xi) &=& \lambda(\sqrt{\frac{\beta}{N}}\omega(\sigma_i z^{\mu}) -
\frac{\beta}{N} \xi_i^{\mu})Z^{\lambda}_{N,p}(\beta;\xi) \\
\label{325}
\partial_{\xi_i^{\mu}} \omega(\sigma_i z^{\mu}) &=&
\sqrt{\frac{\beta}{N}}(\omega((z^{\mu})^{2})-\omega^2(\sigma_i
z^{\mu})) \\ \label{326} \omega((z^{\mu})^2) &=&
Z^{-1}(\beta;\xi)\sum_{\sigma}\int (\prod_{\mu=1}^p
d\mu(z^{\mu}))\frac{\partial}{\partial z^{\mu}} \nonumber
\\&& \Big(z^{\mu}\exp\Big({\sum_{\mu=1}^p \sum_{i=1}^N
(\sqrt{\frac{\beta}{N}}\xi_i^{\mu}\sigma_i z^{\mu} -
\beta\frac{(\xi_i^{\mu})^2}{2N})}\Big)\Big)= \\
&=& 1 + \sqrt{\frac{\beta}{N}}\sum_{i=1}^N
\xi_i^{\mu}\omega(\sigma_i z^{\mu})
\end{eqnarray}
where, in (\ref{324}), $\lambda \in \mathbb{R}$ is a generic positive real
number.

\section{The main results}

Now we are ready to state the main results of this paper, in the form of the following Theorems.

\begin{theorem}\label{intensive}
There is a $\beta_{2}(\alpha)$ defined in the following, such that for $\beta<\beta_{2}(\alpha)$ we have the following limits for the intensive free energy, internal energy and entropy, as $N\to\infty$ and $p/N\to\alpha$:
$$ \lim_{N\to\infty}(-\beta f_{N,p}(\beta;\xi))=\lim_{N\to\infty}N^{-1}\ln Z_{N,p}(\beta;\xi)=\ln 2 -\frac{\alpha}{2}\ln (1-\beta)-\frac{\alpha\beta}{2},$$
$$ \lim_{N\to\infty}(u_{N,p}(\beta;\xi))=-\lim_{N\to\infty}N^{-1}\partial_{\beta}\ln Z_{N,p}(\beta;\xi)=-\frac{\alpha\beta}{2(1-\beta)},$$
$$\lim_{N\to\infty}(s_{N,p}(\beta;\xi))=\lim_{N\to\infty}N^{-1}(\ln Z_{N,p}(\beta;\xi)-\beta \partial_{\beta}\ln Z_{N,p}(\beta;\xi))=\ln 2 -\frac{\alpha}{2}\ln (1-\beta)-\frac{\alpha\beta}{2}-
\frac{\alpha\beta^{2}}{2(1-\beta)},$$
$\xi$-almost surely.
The same limits hold for the quenched averages, so that in particular
$$\lim_{N\to\infty}N^{-1}\mathbb{E}\ln Z_{N,p}(\beta;\xi)=\ln 2 -\frac{\alpha}{2}\ln (1-\beta)-\frac{\alpha\beta}{2}.$$
\end{theorem}

\begin{theorem}\label{fluctuations}
There is a $\beta_{4}(\alpha)$ defined in the following, such that for $\beta<\beta_{4}(\alpha)$ we have the convergence in distribution
\be \ln {\tilde Z}_{N,p}(\beta; \xi)-\ln \mathbb{E}{\tilde Z}_{N,p}(\beta; \xi)\rightarrow C(\beta) + \chi S(\beta) 
\ee 
where $\chi$ is a unit Gaussian in $\mathcal{N}[0,1]$ and
\begin{eqnarray}
C(\beta) &=& -\frac{1}{2}\ln\sqrt{1/(1-\sigma^2\beta^2\alpha)} \\
S(\beta) &=& \Big( \ln \sqrt{1/(1-\sigma^2\beta^2\alpha)}\Big)^{\frac{1}{2}},
\end{eqnarray}
with $\sigma=(1-\beta)^{-1}$.

If we consider the overlaps among $s$ replicas, then there is a $\beta_{2s}(\alpha)$ defined in the following, such that for $\beta<\beta_{2s}(\alpha)$  the rescaled overlaps converge in distribution under $\langle . \rangle = \mathbb{E}\Omega(.)$ as follows
\begin{eqnarray}
\sqrt{N}Q_{ab} &\rightarrow& \frac{\xi_{ab}}{\sqrt{1-\sigma^2\beta^2\alpha}}, \\
\sqrt{p}P_{ab} &\rightarrow&
\frac{\sqrt{\alpha}\beta}{1-\beta^2}\frac{\xi_{ab}}{\sqrt{1-\sigma^2\beta^2\alpha}}+
\frac{1}{1-\beta}\chi_{ab}
\end{eqnarray}
where $\chi_{ab}$ and $\xi_{ab}$ are unit Gaussian in $\mathcal{N}[0,1]$, independent for each couple of replicas $(a,b)$, and independent from the $\chi$ appearing in the limit for the fluctuation of $\ln Z_{N,p}$.
\end{theorem}
 
We remark that the limitation on the parameter regions is strictly related to our technique in the proof. There are good reasons to believe that the theorems can be extended to the whole expected ergodic region $\beta(1+\sqrt{\alpha})<1$.

In order to prove these results, we need some properties about the annealed momenta of the partition function, that will be studied in the next section. 

\section{Momenta of the normalized partition function}

Annealing is a regime in which no retrieval is achievable because
it is implicitly assumed the same time-scale both for neurons and
synapses (the synaptic plasticity is thought of as fast as the
neuronal current rearrangement). Anyway, annealing is the
first step to be investigated in order to have a good control of the statistical mechanics features of the model.
\newline
To obtain the annealed intensive pressure, that we call
$\bar{A}_{N,p}(\beta)$, we must exchange the
logarithm and the average over the memories in the expression of
the quenched free energy (i.e. $\mathbb{E}\ln Z_{N,p}(\beta;\xi)
\Rightarrow \ln \mathbb{E} Z_{N,p}(\beta;\xi)$). Therefore at first we need to
evaluate $\mathbb{E}[Z_{N,p}(\beta;\xi)]$.
\begin{proposition}\label{annealed}
For $0\le\beta <1$ we have that
 \be\label{la22} \mathbb{E}\tilde{Z}_{N,p}(\beta;\xi) = 2^N (1-\beta)^{(-\frac{p}{2})}.\ee
\end{proposition}
\textbf{Proof}
\begin{eqnarray} \nonumber
\mathbb{E}\tilde{Z}_{N,p}(\beta;\xi) &=& \int (\prod_{\mu=1}^p \prod_{i=1}^N d
\mu(\xi_i^{\mu})) \int \prod_{\mu=1}^p d \mu (z_{\mu})
\sum_{\sigma}\exp(\sum_{\mu=1}^p
\sum_{i=1}^N(\sqrt{\frac{\beta}{N}} \xi_i^{\mu}\sigma_iz^{\mu})) \\  &=& \nonumber \sum_{\sigma} \int
(\prod_{\mu=1}^p d\mu(z_{\mu}))\exp((\beta/2)\sum_{\mu=1}^p
z_{\mu}^{^2})=
2^N(1-\beta)^{-\frac{p}{2}}.
\end{eqnarray}
Then, by recalling  the definition (\ref{Ztilde}) of $\tilde{Z}_{N,p}$, we can immediately state the
following
\begin{proposition}\label{Cannfree}
In the thermodynamic limit, and for every value  $0\le\beta<1$,
the annealed pressure per site $\bar{A}(\alpha,\beta) = \lim_{N
\rightarrow \infty}\bar{A}_{N,p}(\beta)$ of the analogical
Hopfield model is
\be\label{annurka} \lim_{N\rightarrow \infty}
 \frac{1}{N}\ln\mathbb{E}({Z}_{N,p}(\beta;\xi)) = \bar{A}(\alpha,\beta)= \ln 2
- \frac{\alpha}{2}\ln(1-\beta)-\frac{\alpha\beta}{2}. \ee
\end{proposition}

\begin{remark}
We stress that the annealed free energy (\ref{annurka}) turns out
to be the same as in the AGS theory with digital patterns. Furthermore,
\be
\lim_{\alpha \rightarrow \infty, \beta \rightarrow 0, \sqrt{\alpha}\beta \rightarrow \beta'} \bar{A}(\alpha,\beta) = \ln 2 + (\beta')^2/4,
\ee
that is the expression for the annealed free
energy of the Sherrington-Kirkpatrick model at the temperature
$\beta'$ \cite{ALR}.
\end{remark}
This is not surprising as, at given $N$, in the limit
$p\rightarrow \infty,\ \beta \rightarrow 0$ with $\beta\sqrt{p}/N \rightarrow \beta'$ we get in distribution $Z_{N,p}(\beta;\xi)\rightarrow Z_N^{SK}(\beta';J)$, where $SK$ stands for the Sherrington-Kirkpatrick model and $J$ is the associated noise (standard unit Gaussian $J_{ij}$ for each couple of sites). At given $N$, the neural network with infinite stored patterns becomes a Sherrington-Kirkpatrick mean field spin glass.
\begin{remark}
If we now turn to the energy density it is easy to show that its
``annealed expression'', defined as $\bar{u}(\alpha,\beta)=- \partial_{\beta}\bar{A}(\alpha,\beta)$, is  given by
\be\label{annU} \lim_{N \rightarrow \infty}
\bar{u}_{N,p}(\beta)
=- \lim_{N \rightarrow \infty} \frac{1}{N}\partial_{\beta}\ln
\mathbb{E}Z_{N,p} = -\frac{1}{2} \frac{\alpha \beta}{(1-\beta)},\ee
where, with respect to (\ref{acca}) thought in the infinite volume limit, the order parameters are missing.
\end{remark}

Both the expressions (\ref{annurka},\ref{annU}) do diverge in the
limit of $\beta \rightarrow 1$ suggesting a point where annealing
has to break, whatever $\alpha$.
\newline
To investigate the region of validity of the annealing, which is defined as the ergodic region, we have to
study the momenta of the partition functions and check if and
where they are well defined.
\newline
It will be
easier to deal with the normalized partition function
$\bar{Z}_{N,p}(\beta;\xi)$ defined as \be\label{tilda}
\bar{Z}_{N,p}(\beta;\xi) \equiv
\frac{\tilde{Z}_{N,p}(\beta;\xi)}{\mathbb{E}\tilde{Z}_{N,p}(\beta;\xi)}. \ee
As the momenta will turn out to be
expressed in terms of overlaps, the following Lemma will be of
precious help for our purpose.
\begin{lemma} At $\beta=0$, 
in the thermodynamic limit the two replica overlaps $q_{ab}$, $a\ne b$, 
become almost surely zero and their rescaled values
$\sqrt{N}q_{ab}$ converge in distribution to unit Gaussian $\xi_{ab}$, independent for each couple of replicas.
\end{lemma}

\textbf{Proof}
\newline
At $\beta=0$ the Boltzmann measure $\Omega_0$ becomes flat on all the configurations, so that all $\sigma^{a}_{i}$'s are independent and  take the values $\pm1$ with equal probability $1/2$. By taking into account the expression of the overlap
$q_{ab}=N^{-1}\sum_i\sigma_i^a\sigma_i^b$, the Lemma follows by the strong law of large numbers and the central limit theorem on sums of independent variables. $\square$
\begin{theorem}\label{Zbarras}
For $s \in \mathbb{N}$ and for  $\beta < \beta_s(\alpha)$, with $\beta_s(\alpha)$ suitably defined in the following,  the following limiting expression for the integer momenta of the normalized partition function holds
\be\label{limZs} \lim_{N \rightarrow
\infty}\mathbb{E}[\bar{Z}_{N,p}^s(\beta;\xi)]=
\exp \Big(\frac{s(s-1)}{4}(\ln(\frac{1}{1-\sigma^2\beta^2\alpha}))\Big),
\ee where $\sigma = 1/(1-\beta)$.
\end{theorem}
\textbf{Proof}
\newline
Let us at first notice that there is a constraint $\beta_s(\alpha)\le s^{-1}$, which becomes effective starting from $s=2$. The key point is that $\mathbb{E}[\tilde{Z}_{N,p}^s(\beta;\xi)]$ becomes infinite, at fixed $N$, if the constraint is not satisfied. In fact, let us calculate this momentum, by introducing the replicated $\sigma^{a}_{i}$, $z^{a}_{\mu}$, for $i=1,2,...,N$, $\mu=1,2,...,p$, $a=1,2,...,s$. 
\begin{eqnarray}
\mathbb{E}\tilde{Z}_{N,p}^s(\beta;\xi) &=&
\mathbb{E}(\sum_{\sigma^1}...\sum_{\sigma^{s}}\int (\prod_{a=1}^{s}\prod_{\mu=1}^p
d \mu(z_{\mu}^{a}))\exp\Big\{\sum_{\mu=1}^p\sum_{i=1}^N \Big(
\sqrt{\frac{\beta}{N}}\xi_i^{\mu}\sum_{a=1}^s
\sigma_i^a z_{\mu}^{a}\Big)\Big\}) \nonumber \\
&=&
\sum_{\sigma^1}...\sum_{\sigma^{s}} \int (\prod_{a=1}^{s}\prod_{\mu=1}^p
d \mu(z_{\mu}^{a}))\exp \Big( \frac{\beta
}{2N}\sum_{\mu=1}^p\sum_{i=1}^N(\sum_{a=1}^s\sigma_i^a z_{\mu}^{a})^2\Big)\nonumber
\\  
&=&
\sum_{\sigma^1}...\sum_{\sigma^{s}} \int (\prod_{\mu=1}^p)\Big((\prod_{a=1}^{s}
d \mu(z_{\mu}^{a}))\exp ( \frac{\beta
}{2N}\sum_{i=1}^N(\sum_{a,b}^s\sigma_i^a z_{\mu}^{a}\sigma_i^b z_{\mu}^{b})\Big),\label{factor}
\end{eqnarray}
where we have performed the Gaussian integration over the memories $\xi$, and have expressed the resulting square in the form of the double sum $\sum_{a,b}$. The sum over $i$ gives the overlaps $q_{ab}$. We notice also that we have complete factorization over $\mu$. Therefore, we can introduce generic variables $z^{a}$, $a=1,2,...,s$, and define
\be\label{B}
B_{p,s}(\beta; Q)=\int (\prod_{a=1}^{s}
d \mu(z^{a}))\exp ( \frac{\beta
}{2}\sum_{a,b}^s q_{ab} z^{a} z^{b}),
\ee
where $Q$ is the $s\times s$ matrix  with elements $q_{ab}$, so that
\be\label{eallap}
\mathbb{E}\tilde{Z}_{N,p}^s(\beta;\xi)=2^{Ns}\Omega_{0}(e^{p \ln B_{p,s}(\beta; Q)}).
\ee
Now we can see the reason for the limitation $\beta<1/s$. In fact, let us notice that 
$$0\le\sum_{a,b}^s q_{ab} z^{a} z^{b}\le
\sum_{a,b}^s |z^{a}||z^{b}|=s^{2}(s^{-1}\sum_{a}|z^{a}|)^{2}\le
s^{2}s^{-1}\sum_{a}(z^{a})^{2}=s\sum_{a}(z^{a})^{2},$$  
where we have introduced the uniform distribution $s^{-1}\sum_{a}$ on $(1,2,...,s)$, and exploited Schwartz inequality. Therefore, the integral defining $B_{p,s}(\beta; Q)$ in (\ref{B}) can be uniformly bound by 
$(1-s\beta)^{-p/2}$, which is finite in the region $s\beta<1$. On the other hand, it is easily seen that for some $\sigma$ configurations the integral in (\ref{B}) is infinite if $s\beta\ge1$. The worst case is when all $q_{ab}=1$. Then we have from (\ref{B})
\begin{eqnarray}\nonumber
B_{p,s}&=&\int (\prod_{a=1}^{s}
d \mu(z^{a}))\exp ( \frac{\beta
}{2}\sum_{a,b}^s z^{a} z^{b})=\int (\prod_{a=1}^{s}
d \mu(z^{a}))\exp ( \frac{\beta
}{2}(\sum_{a} z^{a})^{2})\\ \nonumber &=&\int (\prod_{a=1}^{s}
d \mu(z^{a}))\int d \mu(y) \exp ( \sqrt{\beta}
\sum_{a}z^{a} y)=\int d \mu(y)\exp (\frac{s\beta}{2}y^{2}),
\end{eqnarray}
and the integral on the auxiliary variable $y$ is divergent if $s\beta\ge1$.
From now on we remain in the region $s\beta<1$. Let us go back to the definition (\ref{B}). Write $\sum_{a,b}q_{ab}z^{a}z^{b}=2 \sum_{(ab)}q_{ab}z^{a}z^{b}+ \sum_{a}(z^{a})^{2}$, where $(ab)$ are all couples of different replicas. Then we have
\begin{eqnarray}\nonumber
B_{p,s}(\beta; Q)&=&\int (\prod_{a=1}^{s}(
d \mu(z^{a})e^{\frac12\beta(z^{a})^{2}}))\exp ( \beta
\sum_{(ab)}^s q_{ab} z^{a} z^{b})\\ \nonumber &=&(\int 
d \mu(z^{\prime})e^{\frac12 \beta (z^{\prime})^{2}})^{s}\int (\prod_{a=1}^{s}
d \bar\mu(z^{a}))\exp ( \beta
\sum_{(ab)}^s q_{ab} z^{a} z^{b})\\ &=&(1-\beta)^{-\frac{s}{2}}
\int(\prod_{a=1}^{s}
d \bar\mu(z^{a}))\exp ( \beta
\sum_{(ab)}^s q_{ab} z^{a} z^{b}),\label{B'}
\end{eqnarray}
where we have introduced the normalized deformed measure
$$d \bar\mu(z)= \frac{e^{\frac12 \beta z^{2}}d \mu(z)}{\int e^{\frac12 \beta (z^{\prime})^{2}}d \mu(z^{\prime})}.$$
Finally, if we define a modified $\bar B_{p,s}(\beta; Q)$ as 
\be\label{Bbar}
\bar B_{p,s}(\beta; Q)=\int (\prod_{a=1}^{s}
d \bar\mu(z^{a}))\exp ( \beta
\sum_{(ab)}^s q_{ab} z^{a} z^{b}),\ee
we can write
\be\label{Eallapbar}
\mathbb{E}\tilde{Z}_{N,p}^s(\beta;\xi)=(\mathbb{E}\tilde{Z}_{N,p}(\beta;\xi))^{s}\Omega_{0}(e^{p \ln \bar B_{p,s}(\beta; Q)}).
\ee
In order to investigate the $N\to\infty$ limit, it is convenient to start from the case where $s=2$. Then $\bar B_{p,2}$ can be explicitly calculated in the form 
\be\label{Bbar2}
\bar B_{p,2}=-\frac12\ln (1-\beta^{2}\sigma^{2}q_{12}).
\ee
In fact, in this case we have 
$$\bar B_{p,2}=\int 
d \bar\mu(z^{1}) d \bar\mu(z^{2})\exp ( \beta
 q_{12} z^{1} z^{2}),$$
where the two Gaussian integrals can be calculated explicitly and lead to (\ref{Bbar2}). Therefore, we are led to consider the $(\alpha, \beta)$ region where
$$\Omega_{0}(\exp(-\frac12 p \ln (1-\beta^{2}\sigma^{2}q_{12}^{2})))$$
stays finite in the $N\to\infty$ limit. Through a simple change of variables $\sigma_{i}^{1}=\sigma_{i}^{\prime}, \sigma_{i}^{2}=\sigma_{i}^{\prime} \sigma_{i}$ we are led to the consideration of a mean field ferromagnetic Ising system $(\sigma_{1},...,\sigma_{N})$  with normalized partition function 
\be\label{Omega}\Omega_{0}(\exp(-\frac12 p \ln (1-\beta^{2}\sigma^{2}m^{2}))),\ee
where now $\Omega_{0}=2^{-N}\sum_{\sigma}$, and $m=N^{-1}\sum_{i}\sigma_{i}$, as usual.
Now, we can state and prove the following Theorem.
\begin{theorem}\label{beta2}
Consider the trial function
$$\phi(\alpha,\beta; M)=-\frac12 \alpha \ln (1-\beta^{2}\sigma^{2}M^{2})+\ln\cosh (\alpha \frac{\beta^{2}\sigma^{2}M}{1-\beta^{2}\sigma^{2}M^{2}})-\alpha M \frac{\beta^{2}\sigma^{2}M}{1-\beta^{2}\sigma^{2}M^{2}},$$
depending on the order parameter $M$, with $-1\le M\le 1$. Clearly, at $M=0$ we have $\phi(\alpha,\beta; 0)=0$. Define $\beta_{2}(\alpha)$ as the largest value such that, for any $\beta<\beta_{2}(\alpha)$, we have $\phi(\alpha,\beta; M)<0$ for any positive $M$. It is easily shown that $\beta_{2}(\alpha)\ge (1+\sqrt{1+\alpha})^{-1}$. Then for $\beta<\beta_{2}(\alpha)$ we have 
\be\label{limOmega0}\lim_{N\to\infty}\Omega_{0}(\exp(-\frac12 p \ln (1-\beta^{2}\sigma^{2}m^{2})))= (1-\alpha \beta^{2}\sigma^{2})^{-\frac12}.\ee
\end{theorem}
Notice that $\beta_{2}(\alpha)$ defines the onset of the ferromagnetic phase transition for the model. The proof of the Theorem follows standard methods of statistical mechanics. In order to get a lower bound we only notice that
$$-\ln (1-\beta^{2}\sigma^{2}m^{2})\ge \beta^{2}\sigma^{2}m^{2},$$
and therefore
$$\Omega_{0}(\exp(-\frac12 p \ln (1-\beta^{2}\sigma^{2}m^{2})))\ge \Omega_{0}(\exp(\frac12 p \beta^{2}\sigma^{2}m^{2})).$$      
The r.h.s. converges to $(1-\alpha \beta^{2}\sigma^{2})^{-\frac12}$, provided $\alpha \beta^{2}\sigma^{2}<1$. Notice that this last condition can be written also as $\beta<1/(1+\sqrt{\alpha})$, which is the critical line according to the AGS theory. For the upper bound, let us introduce the truth functions on the $\sigma$ configuration space, defined by $\chi_{1}=\chi(m^{2}\le \bar m^{2})$, and $\chi_{2}=\chi(m^{2}> \bar m^{2})$, where $\bar m$ is some positive number. Then we have that the $\Omega_{0}(...)$ in (\ref{Omega}) splits into the sum of two pieces
$$\Omega_{0}(...)=\Omega_{0}(...\chi_{1})+\Omega_{0}(...\chi_{2}).$$ For $\beta<\beta_{2}(\alpha)$, the second piece converges to zero as $N\to\infty$. In fact, let us define for $-1\le M\le 1$
\be \label{psi}
\psi(M)=-\frac{p}{2N}\ln (1-\beta^{2}\sigma^{2}M^{2}),\ee
with its $M$ derivative
\be\label{psi'}
\psi^{\prime}(M)=\frac{p}{N}\frac{\beta^{2}\sigma^{2}M}{1-\beta^{2}\sigma^{2}M^{2}}.\ee
Notice that $\psi$ is convex in $M$, so that
\be\label{convex}
\psi(m) \ge \psi(M) + (m-M)\psi^{\prime}(M).\ee
Now, let us consider $M$ as taking all values of $m$ for which $m^{2}> \bar m^{2}$. There are at most $N+1$ such values. For the sake of simplicity, introduce the inequality
\be
1\le \sum_{M}\exp\Big(-N\big(\psi(m)-\psi(M)-(m-M)\psi^{\prime}(M)\big)\Big),\ee
where $M$ is summed over all stated values. The reason of the inequality is clear. In fact, there is one term equal to $1$, when $M=m$, while all other terms are positive. By exploiting the inequality, we can write
\be
\Omega_{0}(\exp(-\frac12 p \ln (1-\beta^{2}\sigma^{2}m^{2}))\chi_{2})\le \sum_{M}\Omega_{0}(\exp\Big(-N\big(-\psi(M)-(m-M)\psi^{\prime}(M)\big)\Big)\chi_{2}).\ee
If now we remove the $\chi_{2}$, and perform the average over $\Omega_{0}$, by taking into account that the exponent is factorized with respect to the $\sigma_{i}$'s, we get
\be
\Omega_{0}(\exp(-\frac12 p \ln (1-\beta^{2}\sigma^{2}m^{2}))\chi_{2})\le \sum_{M}\exp\Big(N\big(\psi(M)+\ln \cosh (\frac{p}{N}\frac{\beta^{2}\sigma^{2}M}{1-\beta^{2}\sigma^{2}M^{2}})-M\psi^{\prime}(M)\big)\Big).\ee
Clearly, in the region $\beta<\beta_{2}(\alpha)$ and for large $N$, each term in the sum is uniformly bounded by an exponential of the type $\exp(-c N)$, for some constant $c$. Of course there are at most $N+1$ terms. Therefore, as $N\to\infty$, the r.h.s. goes to zero, as we were interested to show.\newline
Now we must consider the first term $\Omega_{0}(...\chi_{1})$. Let us notice that in the region $m^{2}\le\bar m^{2}$ by convexity we have $-\ln (1-\beta^{2}\sigma^{2}m^{2})\le -\ln (1-\beta^{2}\sigma^{2}\bar m^{2})(m^{2}/\bar m^{2})$. By inserting the inequality in the first term, and neglecting the $\chi_{1}$, we have in the infinite volume limit
\be
\limsup_{N\to\infty}\Omega_{0}(\exp(-\frac12 p \ln (1-\beta^{2}\sigma^{2}m^{2}))\chi_{1})\le (1+\frac{\alpha\ln (1-\beta^{2}\sigma^{2}m^{2})}{\bar m^{2}})^{-\frac12}.\ee
Since $\bar m$ is arbitrary, we can take $\bar m\to0$. Collecting all results, we immediately establish the limit in (\ref{limOmega0}). Notice that $\beta_{2}(\alpha)\le (1+\sqrt{\alpha})^{-1}$, otherwise $\phi(\alpha,\beta; M)$ would start with a positive derivative at $M=0$. This ends the proof of Theorem \ref{Zbarras} in the case $s=2$.\newline
The general case can be handled in a similar way. Now we encounter ferromagnetic models for the Ising variables $\sigma^{a}_{i}, a=1,2,...,s, i=1,2,...,N$ with \textit{Boltzmannfaktor} $\exp(p \ln \bar B_{p,s})$. If $\beta_{s(\alpha)}$ denotes the onset of the associated ferromagnetic transition, then we can immediately prove that
\be
\lim_{N\to\infty} \Omega_{0}(e^{p \ln \bar B_{p,s}})=
\exp \Big(\frac{s(s-1)}{4}(\ln(\frac{1}{1-\sigma^2\beta^2\alpha}))\Big),\ee
for $\beta\le \beta_{s(\alpha)}$.
In fact, as in the proof for the case $s=2$, we see that
in the expression of $\bar B_{p,s}(\beta; Q)$ only the first two terms in the expansion of the exponent do matter, provided the stated condition on $\beta$ holds. These terms are easily calculated as in the case $s=2$. Then we recall that under $\Omega_{0}$, for $a\ne b$, the rescaled overlaps $\sqrt{N}q_{ab}$ converge in distribution to independent unit Gaussian 
$\xi_{ab}$. We see that the term $s(s-1)$ in formula (\ref{limZs}) comes essentially from the fact that there are $s(s-1)/2$ couples $(a,b)$ for $s$ replicas.  
Therefore, Theorem \ref{Zbarras} is fully established. $\Box$    

Now we are ready to prove Theorem \ref{intensive}, at least in the region $\beta<\beta_{2}(\alpha)$.
\newline
First of all, let us recall that if $u_{N}\ge0$ is a sequence of random variables normalized to $\mathbb E (u_{N})=1$, then a simple application of the Markov inequality \cite{Talabook} and the Borel-Cantelli Lemma gives
$$\limsup_{N\to\infty} \frac{1}{N}\ln u_{N} \le 0,$$
almost surely. Moreover, if $\mathbb E (u_{N}^{2})\le c E^{2} (u_{N})$, uniformly in $N$, for some finite constant $c$, then
$$ \lim_{N\to\infty} \frac{1}{N}\ln u_{N} = 0,$$
almost surely. 
\newline
If we define $$u_{N}=\frac{\mathbb E (\tilde Z_{N,p}(\beta;\xi)^{s}}{\mathbb E^{s} (\tilde Z_{N,p}(\beta;\xi)},$$ and take into account our previous results, then we immediately find, $\xi$-almost surely
\begin{eqnarray}
\limsup_{N\to\infty} \frac{1}{N}\ln \tilde Z_{N,p}(\beta;\xi)&\le&
\ln2-\frac{\alpha}{2}\ln(1-\beta),\ \text{for any}\ \beta<1,\\
\lim_{N\to\infty} \frac{1}{N}\ln \tilde Z_{N,p}(\beta;\xi)&=&
\ln2-\frac{\alpha}{2}\ln(1-\beta),\ \text{for any}\ \beta<\beta_{2}(\alpha). 
\end{eqnarray}
In order to get Theorem \ref{intensive}, in the stated region, it is only necessary to recall the equation (\ref{ZZtilde}) connecting $Z_{N,p}$ with $\tilde Z_{N,p}$, and the limiting properties of $\hat f_{N}$. $\Box$ 

\section{Log-normality of the limiting distribution for the partition function}

In this section we want to  show that the limiting distribution of the normalized partition
function (\ref{tilda}), at least in a given region of the $\alpha,\beta$ plane, is log-normal. This will
immediately give us the mean and the fluctuations of the
thermodynamical quantities in that region \cite{moment}.
\newline
Let us remember that if $C(\beta)$ and $S(\beta)$ are given
functions and $\chi$ a standard gaussian $\mathcal{N}[0,1]$, a
variable $\eta(\beta)$ has a log-normal distribution if it is
possible to write it down as \cite{ellis}
\be \eta(\beta)=\exp(C(\beta)+\chi S(\beta)). \ee 
\newline
The momenta of $\eta(\beta)$ are \be
\mathbb{E}[\eta^s(\beta)]=\mathbb{E}[\exp(s C(\beta) + s
S(\beta)\chi) ]= \exp(C(\beta) s + \frac{1}{2}s^2 S^2(\beta)). \ee
So if we choose
\begin{eqnarray}
C(\beta) &=&
-\frac{1}{2}\ln(\sqrt{\frac{1}{1-\sigma^2\beta^2\alpha}}) \\ 
S^2(\beta) &=& \ln(\sqrt{\frac{1}{1-\sigma^2\beta^2\alpha}}) 
\end{eqnarray}
we see that $\bar{Z}_{N,p}(\beta;\xi)$ and $\eta(\beta)$ have the same
integer momenta in the limit, provided we restrict the values of $\beta$, according to the order of the momentum $s$, as expressed in Theorem \ref{Zbarras}. This seems to suggest that $\bar{Z}_{N,p}(\beta;\xi)$ is
log-normal distributed in the limit. To prove that this is effectively the case it will be sufficient to prove that the momenta  $\bar{Z}_{N,p}(\beta;\xi)^{\lambda}$ do in fact converge to those of $\eta$ for all values of $\lambda$ in some interval of the real line \cite{carle}. In other words, we have to extend the limiting behavior of Theorem \ref{Zbarras}, from integer values of $s$ to real values $\lambda$ in some nontrivial interval, at least in some region of the $(\alpha,\beta)$ plane.
To solve this task we have to evaluate the limiting behavior of 
$\bar{Z}_{N,p}^{\lambda}(\beta;\xi)$, for $\lambda$ in some interval of the real line. 
\newline
We get the result by analyzing \be\label{371}
\partial_{\beta}\ln\mathbb{E}[\bar{Z}_{N,p}^{\lambda}] =
\partial_{\beta}(\ln \mathbb{E}[\tilde{Z}_{N,p}^{\lambda}]-\lambda \ln
\mathbb{E}[\tilde{Z}_{N,p}])=
\frac{\partial_{\beta}\mathbb{E}[\tilde{Z}_{N,p}^{\lambda}]}{\mathbb{E}[\tilde{Z}_{N,p}^{\lambda}]}
- \lambda\frac{\partial_{\beta}\mathbb{E}[\tilde{Z}_{N,p}]}{\mathbb{E}[\tilde{Z}_{N,p}]},
\ee that can be written in terms of overlaps via the
following helpful Proposition.
\begin{proposition}\label{dbeta}
For any real $\lambda$, with $\lambda\le s$, $s$ integer, and $\beta<1/s$, the $\beta$-derivative of the annealed real momenta of the
partition function can be expressed in terms of overlaps as 
\be\label{dbetaE}
\frac{\partial_{\beta}\mathbb{E}[\bar{Z}_{N,p}^{\lambda}]}{\mathbb{E}[\bar{Z}_{N,p}^{\lambda}}=
\frac{p \lambda / 2 }{\big( 1 - \beta   \big)}
\left((\lambda-1)\mathbb{E}\left(\frac{\bar{Z}^{\lambda}_{N,p}]}{\mathbb{E}[\bar{Z}_{N,p}^{\lambda}]}\Omega(q_{12}p_{12})
\right)+1\right). \ee
\end{proposition}
Notice that $\partial_{\beta}\ln\mathbb{E}[\bar{Z}_{N,p}^{\lambda}]$ is convex increasing in $\lambda$. Therefore, the limitation on the values of $\beta$ assures the existence of the relevant averages.\newline
\textbf{Proof}
\newline
Using equation (\ref{323}) we can write
\be\label{373}
\partial_{\beta}\mathbb{E}[\tilde{Z}_{N,p}^{\lambda}] = \mathbb{E} [\lambda
\tilde{Z}_{N,p}^{\lambda-1}\partial_{\beta}\tilde{Z}_{N,p} ]= \sum_{\mu=1}^p\sum_{i=1}^N
\Big( \frac{\lambda}{2\sqrt{\beta
N}}\mathbb{E}[\xi_i^{\mu}Z_{N,p}^{\lambda}\omega(\sigma_iz^{\mu})]\Big). \ee
Furthermore we can write
\begin{eqnarray}
\mathbb{E}[\xi_i^{\mu}\tilde{Z}_{N,p}^{\lambda}\omega(\sigma_i z^{\mu})] &=&
\lambda \sqrt{\frac{\beta}{N}}\mathbb{E}[\tilde{Z}_{N,p}^{\lambda}\omega^2(\sigma_i
z^{\mu})] - \lambda
\frac{\beta}{N}\mathbb{E}[\xi_i^{\mu}\tilde{Z}_{N,p}^{\lambda}\omega(\sigma_i
z^{\mu})]  \\  \nonumber &+&
\sqrt{\frac{\beta}{N}}\mathbb{E}[\tilde{Z}_{N,p}^{\lambda}] +
\frac{\beta}{N}\sum_{j=1}^N\mathbb{E}[\tilde{Z}_{N,p}^{\lambda}\xi_j^{\mu}\omega(\sigma_iz^{\mu})]
- \sqrt{\frac{\beta}{N}}\mathbb{E}[\tilde{Z}_{N,p}^{\lambda}\omega^2(\sigma_i
z^{\mu})],
\end{eqnarray}
and, using (\ref{324},\ref{325},\ref{326}), \be\label{377}
\sum_{\mu=1}^p\sum_{i=1}^N
\mathbb{E}[\xi_i^{\mu}\tilde{Z}_{N,p}^{\lambda}\omega(\sigma_i z^{\mu})] =
\frac{p\sqrt{\beta N}}{(1-\beta+\beta\lambda/N)}\Big( (\lambda-1)
\mathbb{E}[\tilde{Z}_{N,p}^{\lambda}\Omega(q_{12}p_{12})] +
\mathbb{E}[\tilde{Z}_{N,p}^{\lambda}] \Big). \ee By substituting (\ref{377})
into (\ref{373}) and dividing by $\mathbb{E}(Z_{N,p}^{\lambda})$ we
get the result. \ \ \ \ $\Box$
\newline
\newline
\newline
With the help of  (\ref{dbetaE}) we can rewrite
(\ref{371}) as
\be\label{378}
\partial_{\beta}\ln\mathbb{E}[\bar{Z}_{N,p}^{\lambda}] =
\frac{\sqrt{\alpha}\lambda(\lambda-1)}{2(1-\beta)}\mathbb{E}\Big( \frac{\bar{Z}_{N,p}^{\lambda}}{\mathbb{E}(\bar{Z}_{N,p}^{\lambda})} \Omega(\sqrt{N}q_{12}\sqrt{p}p_{12}) \Big).
\ee
To proceed further we have now to investigate the distribution of
the rescaled overlaps because they  do appear into the expression
above. Such distribution can be obtained trough the evaluation of
their momenta generating function. We will see that at least in a given region the distribution of $\bar{Z}_{N,p}^{\lambda}$ is not
coupled with the one of the overlaps. From this observation, by looking at  eq. (\ref{382}) the log-normality for the normalized partition function is easily achieved.
\newline
Let us start by proving the following
\begin{proposition}\label{teorema1}
Consider a generic number of replicas $s$. For each couple $(a,b)$ of replicas, let $(\lambda_{ab}, \eta_{ab})$ be real numbers in the momentum generating functional (which we assume to be very small). Let $\lambda$ be a real number in the interval $s\le\lambda\le 2s$. Then, at least for $\beta<\beta_{2s}(\alpha)$, we have the limit 
\begin{eqnarray}\label{382}
&& \lim_{N \rightarrow \infty}\mathbb{E} \Big(
\frac{\bar{Z}_{N,p}^{\lambda}}{\mathbb{E}[\bar{Z}_{N,p}^{\lambda}]}\Omega\Big(
\exp(\sum_{ab}\lambda_{ab}\sqrt{N}q_{ab} +
\sum_{ab}\eta_{ab}\sqrt{p}p_{ab}) \Big)\Big) \\ \nonumber && =
\exp\Big( \frac{1}{2}\sum_{ab}\lambda_{ab}^2
\frac{1}{1-\sigma^2\beta^2\alpha} +
\frac{1}{2}\sum_{ab}\eta_{ab}^2\sigma^2(\frac{\alpha\beta^2\sigma^2}{1-\sigma^2\beta^2\alpha}
+ 1) +
\sum_{ab}\frac{\sqrt{\alpha}\beta\sigma^2\lambda_{ab}\eta_{ab}}{1-\sigma^2\beta^2\alpha}
\Big)
\end{eqnarray}
where as usual $\sigma=1/(1-\beta)$ and
$\sum_{ab}$ denotes the sum over all \textit{couples} $(ab)$.
\end{proposition}
\textbf{Proof}
\newline
We give the proof at first for  $\lambda=s$, in the region $\beta<\beta_{s}(\alpha)$. Then we will enlarge
the proof to the interval $s\le\lambda\le 2s$, in the region $\beta<\beta_{2s}(\alpha)$. For $\lambda=s$ the l.h.s.
of (\ref{382}) can be thought of as
\begin{eqnarray}\nonumber
&& (\prod_{ia}\sum_{\sigma^{a}_{i}}) \int (\prod_{\mu a} d \mu(z_{\mu}^a))
e^{\Big(\frac{\beta}{2}\sum_{\mu=1}^p\sum_{ab}q_{ab}z_{\mu}^{a}z_{\mu}^{b}\Big)}
e^{\Big(\sum_{ab}\lambda_{ab}(\sqrt{N}q_{ab})
+ \sum_{ab}\eta_{ab}\sqrt{p}p_{ab} \Big)}
=
\\ \nonumber
&& (\prod_{ia}\sum_{\sigma^{a}_{i}}) \int (\prod_{\mu a} d \mu_{\sigma}(z_{\mu}^a))\exp \Big(
\sum_{ab}\sqrt{p} p_{ab}(\sqrt{\alpha}\beta \xi_{ab} +
\eta_{ab})
\Big)\exp(\sum_{ab}\lambda_{ab}\xi_{ab})\sigma^{\frac{ps}{2}}
= \\ \nonumber && \Omega_0 \Big(
\exp{(\sum_{ab}\frac{1}{2}\sigma^2(\alpha \beta^2
\xi_{ab}^2 + \eta^2_{ab} + 2\sqrt{\alpha} \beta
\eta_{ab}\xi_{ab}) )}
\Big)\exp(\sum_{ab}\lambda_{ab}\xi_{ab})\sigma^{\frac{ps}{2}}2^{Ns}
= \\ \nonumber && \exp\Big( \frac{1}{2}\sum_{ab}
\frac{1}{1-\sigma^2\beta^2\alpha}(\alpha\beta^2
\sigma^4 \eta_{ab}^2 + \lambda_{ab}^2 + 2\sqrt{\alpha}\beta
\sigma^2\lambda_{ab}\eta_{ab}) +
\frac{\sigma^2}{2}\eta_{ab}^2 \Big) \mathbb{E}[Z_{N,p}^s],
\end{eqnarray}
where $d\mu_{\sigma}$ is the Gaussian with variance $\sigma=(1-\beta)^{-1}$.
Therefore, by taking the limit, we prove the proposition $\lambda=s$. In order to provide the extension to the interval $s\le\lambda\le 2s$ we must show that defining
\be\nonumber \mathcal{A} \equiv \exp\Big(
\frac{1}{2}\sum_{ab}\lambda_{ab}^2
\frac{1}{1-\sigma^2\beta^2\alpha}+
\frac{1}{2}\sum_{ab}\eta_{ab}^2\sigma^2(\frac{\alpha \beta^2
\sigma^2}{1-\sigma^2\beta^2\alpha}+1)+
\sum_{ab}\frac{\sqrt{\alpha}\beta\sigma^2}{1-\sigma^2\beta^2\alpha}
\lambda_{ab}\eta_{ab} \Big) \ee
the following holds \be\label{863} \lim_{N\to\infty}\mathbb{E}\Big(
\bar{Z}_{N,p}^{\lambda}\Omega(\exp(\sum_{ab}\lambda_{ab}\sqrt{N}q_{ab}
+ \sum_{ab}\eta_{ab}\sqrt{p}p_{ab}) - \mathcal{A}) \Big) = 0. \ee

The proof of (\ref{863}) can be obtained in the simplest way by using
Cauchy-Schwartz  inequality $(\mathbb{E}^2[AB]\leq
\mathbb{E}[A^2]\mathbb{E}[B^2])$, choosing $\lambda = \mu + s$,
with $0\le\mu\le s$. In fact, we have 
\begin{eqnarray}\nonumber
&\mathbb{E}^2&\left( \bar{Z}^{\mu}_{N,p}
\bar{Z}^{s}_{N,p}\left(\Omega(\exp(\sum_{ab}\lambda_{ab}\sqrt{N}q_{ab}
+ \sum_{ab}\eta_{ab}\sqrt{p}p_{ab}))-\mathcal{A}\right)\right) \leq \\
\label{2fac}
&\mathbb{E}&(\bar{Z}_{N,p}^{2\mu})\mathbb{E}\left(\bar{Z}_{N,p}^{2s}\left(\Omega(\exp(\sum_{ab}\lambda_{ab}
\sqrt{N}q_{ab}+\sum_{ab}\eta_{ab}\sqrt{p}p_{ab}))-\mathcal{A}\right)^2\right).
\end{eqnarray}
Here the first factor is bounded in the region $\beta<\beta_{2s}(\alpha)$. In fact, by monotonicity we have $\mathbb{E}(\bar{Z}_{N,p}^{2\mu})\le \mathbb{E}(\bar{Z}_{N,p}^{2s})$.  
The second term is the sum of three terms obtained by calculating
the square of \\ $\left(\Omega(\exp(\sum_{ab}\lambda_{ab}
\sqrt{N}q_{ab}+\sum_{ab}\eta_{ab}\sqrt{p}p_{ab}))-\mathcal{A}\right),$
the simplest being $\mathbb{E}(\bar{Z}_{N,p}^{2s})\mathcal{A}^2$ which is known.
\newline
It is easy to check that for the
double-product we have in the limit \be\nonumber 
\lim_{N\to\infty} \mathcal{A} \mathbb{E}
\left(
\bar{Z}_{N,p}^{2s}\Omega(\exp(\sum_{ab}\lambda_{ab}\sqrt{N}q_{ab}+\sum_{ab}\eta_{ab}\sqrt{p}p_{ab}))\right)
= \lim_{N\to\infty} \mathcal{A}^2\mathbb{E}(\bar{Z}_{N,p}^{2s}).\ee
For the last term we have \begin{eqnarray}\nonumber
\lim_{N\to\infty} &&\mathbb{E}\left(
\bar{Z}_{N,p}^{2s}\Omega(\exp(\sum_{ab}\lambda_{ab}\sqrt{N}q_{ab} +
\sum_{ab}\eta_{ab}\sqrt{p}p_{ab} +
\sum_{\tilde{a}\tilde{b}}\lambda_{\tilde{a}\tilde{b}}\sqrt{N}q_{\tilde{a}\tilde{b}}
+\sum_{\tilde{a}\tilde{b}}\eta_{\tilde{a}\tilde{b}}\sqrt{p}p_{\tilde{a}\tilde{b}}))\right)
 \\ \lim_{N\to\infty} && =\mathcal{A}^2\mathbb{E}(\bar{Z}_{N,p}^{2s}),
\end{eqnarray}
where the state $\Omega$ is thought of by $2s$ replicas, the sum
on the couples of variables $(\tilde{a},\tilde{b})$ taking into account the
second set $s+1,...,2s$.
\newline
The sum of these three terms goes to zero as $N \rightarrow
\infty$ proving the Proposition. $\Box$
\bigskip

From Proposition (\ref{teorema1})  we can derive the next Corollary,  which is a part of Theorem \ref{fluctuations}.
\begin{corollary} For $s$ replicas, at least for $\beta<\beta_{2s}(\alpha)$, 
in the thermodynamic limit the limiting distribution of the
rescaled overlaps  are
\begin{eqnarray}
\sqrt{N}Q_{ab} &\rightarrow& \xi_{ab} \\
\sqrt{p}P_{ab} &\rightarrow&
\frac{\sqrt{\alpha}\beta}{1-\beta^2}\xi_{ab}+
\frac{1}{1-\beta}\chi_{ab}
\end{eqnarray}
where $\chi_{ab} \in \mathcal{N}(0,1)$ and $\xi_{ab} \in
\mathcal{N}(0,1/(1-\sigma^2\beta^2\alpha))$.
\end{corollary}
Now we are ready for the proof of the following basic Theorem. 
\begin{theorem}\label{teorema3}
For $\beta < \beta_4(\alpha)$, in the thermodynamic
limit, we have that in distribution \be \bar{Z}_{N,p}(\beta; \xi)\rightarrow \exp\Big( C(\beta) + \chi S(\beta) \Big)
\ee where $\chi \in \mathcal{N}[0,1]$ and
\begin{eqnarray}
C(\beta) &=& -\frac{1}{2}\ln\sqrt{1/(1-\sigma^2\beta^2\alpha)} \\
S(\beta) &=& \Big( \ln \sqrt{1/(1-\sigma^2\beta^2\alpha)}\Big)^{\frac{1}{2}}.
\end{eqnarray}
\end{theorem}
\textbf{Proof}
\newline
The limitation $\beta < \beta_4(\alpha)$ is clear. In fact, we will exploit formula (\ref{378}), which requires two replicas, for $2\le\lambda\le4$, and the results of Proposition (\ref{teorema1}), that require the stated limitation on $\beta$.  
Therefore, we see immediately that
\begin{eqnarray} &&\lim_{N \rightarrow \infty} \mathbb{E}\Big(
\bar{Z}_{N,p}^{\lambda}\Omega(\sqrt{N}q_{ab}\sqrt{p}p_{ab}) /
\mathbb{E}( \bar{Z}_{N,p}^{\lambda})\Big) \\
&=& \partial_{\lambda_{ab}}\partial_{\eta_{ab}}\lim_{N\rightarrow
\infty}\frac{\mathbb{E}\left(
\bar{Z}_{N,p}^s\Omega(\exp(\sum_{ab}\lambda_{ab}\sqrt{N}q_{ab}+\sum_{ab}\eta_{ab}\sqrt{p}p_{ab}))
\right)}{\mathbb{E}(\bar{Z}_{N,p}^{\lambda})} \\
&=& \frac{\sqrt{\alpha}\beta\sigma^2}{(1-\sigma^2\beta^2\alpha)}.
\end{eqnarray}
Therefore, we have
\be
\lim_{N\to\infty}\partial_{\beta}\ln \mathbb{E}[\bar{Z}^{\lambda}] =
\frac{\lambda(\lambda-1)}{2}\Big( \frac{1}{1-\beta}\frac{\alpha
\beta}{(1-\beta)^2-\alpha\beta^2} \Big). \ee By exploiting convexity in $\beta$, we can integrate this expression and obtain the limit
$$
\lim_{N \rightarrow \infty}
\mathbb{E}[\bar{Z}_{N,p}^{\lambda}]\rightarrow \exp\Big(
\frac{\lambda(\lambda-1)}{4}(\ln(\frac{1}{1-\sigma^2\beta^2\alpha})) \Big),
$$
for all values of $\lambda$ in a nontrivial interval of the real axis. This shows the convergence in distribution of $\bar{Z}_{N,p}$ to a log-normal random variable, as stated in the Theorem.\ \ \ $\Box$
\newline
Finally, we can easily prove Theorem \ref{fluctuations} if we recall the definition in (\ref{tilda}). \ \ \ $\Box$

\section{Conclusion}

In this work we introduced the framework of the real
replicas, successfully applied on spin-glasses (see e.g.
\cite{barra1}\cite{Gsum}\cite{gt2}), to neural networks in the
ergodic regime. This approach naturally holds for the high storage
memory case, which is mathematically challenging. Acting together as
a biological generalization to analogical stored memories and as a
technical trick to manage easily the mathematical control, we
allowed the patterns to live as Gaussian variables on $\mathcal{N}[0,1]$ instead of $\pm
1$ but, as we checked a fortiori, this does not affect (at least
in the part of the ergodic region that we can control) any macroscopical distribution once the
thermodynamic limit is taken. Thinking at the Hopfield model as a
bipartite model in a proper space of variables, beyond the Mattis
magnetization, we introduced the other order parameters $q_{ab}$ and
$p_{ab}$, one for each interacting structure, the $N$ dichotomic
Ising neurons $\sigma_i$ and the $p$ fictitious Gaussians
$z_{\mu}$, which are able to fully describe the high temperature
region we investigated.
\newline
We showed that the  partition function is log-normal
distributed in a suitably defined regionn, then we evaluated the distribution of the rescaled
overlaps, which share centered Gaussian fluctuations with
different variances. Finally we proved that all the
thermodynamic quantities fluctuate around their annealed
approximation and calculated their spread.
All the densities (e.g. energy density, free energy density and entropy density) turn out
to be self-averaging  on their annealed values.
\newline
Further investigation should give us the full control of the whole ergodic phase and  bring us exploring the retrieval phase and hopefully the still completely obscure broken replica
phase.

\begin{acknowledgments}
Support from MiUR (Italian Ministry of
University and Research) and INFN (Italian Institute for Nuclear
Physics) is gratefully acknowledged.
\newline
AB is grateful to Peter Sollich, Alberto Bernacchia and Gianluigi
Mongillo for useful discussions; his work is partially supported by the
SmartLife Project (Ministry Decree $13/03/2007$ n.$368$) and partially by the
 CULTAPTATION Project (European Commission
contract FP6 - 2004-NEST-PATH-043434)
\end{acknowledgments}


\addcontentsline{toc}{chapter}{References}
\begin{thebibliography}{9}


\bibitem{tirozzi1} S. Albeverio, B. Tirozzi, B. Zegarlinski {\em Rigorous results for the free energy in the
Hopfield model}, Comm. Math. Phys. \textbf{150}, 337 (1992).

\bibitem{amit} D.J. Amit, {\em Modeling brain function: The world of attractor neural
network}, \ Cambridge University Press, (1992).

\bibitem{ags1} D.J. Amit, H. Gutfreund, H. Sompolinsky {\em Spin Glass model of neural
networks}, Phys. Rev. A \textbf{32}, 1007-1018,  (1985).

\bibitem{ags2} D.J. Amit, H. Gutfreund, H. Sompolinsky {\em Storing infinite numbers of patterns in a spin glass model of neural networks},
Phys. Rev. Lett. \textbf{55}, 1530-1533,  (1985).


\bibitem{barra1} A. Barra,
\textit{Irreducible free energy expansion and overlap locking in
mean field spin glasses}, J. Stat. Phys. \textbf{123}, 601-614
(2006).

\bibitem{barra2} A. Barra,
\textit{The mean field Ising model trhough interpolating
techniques}, J. Stat. Phys. \textbf{132}, 787-809
(2008).

\bibitem{hotel} A. Bernacchia, D.J.Amit, {\em Impact of spatiotemporally correlated images
on the structure of memory}, Proc. Natl. Acad. Sci. USA, \textbf{104},
3544-3549 (2007).

\bibitem{alb4} A. Bernacchia, P. Naveau, {\em Detecting spatial patterns with the cumulant function: the theory}, Nonlin. Processes Geophys. $15:159$ (2008).

\bibitem{bovier1}A. Bovier, B. Niederhauser, {\em The spin-glass phase-transition in the Hopfield model with p-spin interactions}, Adv. Theor. Math. Phys. \textbf{5}, $1001-1046$ (2001).

\bibitem{bovier2} A. Bovier, A.C.D. van Enter and B. Niederhauser, {\em Stochastic symmetry-breaking in a Gaussian Hopfield-model}, J. Stat. Phys.  \textbf{95}, 181-213 (1999).

\bibitem{bovier3} A. Bovier, V. Gayrard {\em An almost sure central limit theorem for the Hopfield model}, Markov Proc. Rel. Fields  \textbf{3}, 151-173 (1997).

\bibitem{bovier4} A. Bovier {\em Self-averaging in a class of generalized Hopfield models}, J. Phys. A \textbf{27}, 7069-7077 (1994).

\bibitem{carle} N.I.Akhiezer, {\em The Classical
Moment Problem and Some Related Questions in Analysis},
Oliver-Boyd, 1965.


\bibitem{comets} F. Comets, J. Neveu, {\em The Sherrington-Kirkpatrick
model of spin glasses and stochastic calculus: the high
temperature case}, Commun. Math. Phys. {\bf 166}, 549 (1995).

\bibitem{peter} A.C.C. Coolen, R. Kuehn, P. Sollich, {\em Theory of Neural Information Processing
Systems}, Oxford University Press, 2005.

\bibitem{ellis} R.S. Ellis,
{\em Large deviations and statistical mechanics}, Springer, New
York, 1985.

\bibitem{alb2} A. Engel, C. Van den Broeck, {\em Statistical Mechanics of
Learning}, Cambridge University Press, 2001.

\bibitem{guerrasg} F. Guerra, {\em An introduction to mean field spin glass theory: methods and results},
In: \textit{Mathematical Statistical Physics}, A. Bovier et al. eds,
$243-271$, Elsevier, Oxford, Amsterdam, 2006.


\bibitem{broken} F. Guerra, {\em Broken Replica Symmetry Bounds in the
Mean Field Spin Glass Model}, Commun, Math. Phys. \textbf{233:1},
1-12 (2003).

\bibitem{guerra2} F. Guerra, {\em About the overlap distribution in mean field
spin glass models}, Int. Jou. Mod. Phys. B {\bf 10}, 1675-1684
(1996).

\bibitem{Gsum} F. Guerra, {\em Sum rules for the free energy in the mean
field spin glass model}, in {\em Mathematical Physics in
Mathematics and Physics: Quantum and Operator Algebraic Aspects},
Fields Institute Communications {\bf 30}, American Mathematical
Society (2001).

\bibitem{limterm} F. Guerra, F. L. Toninelli, {\em
The Thermodynamic Limit in Mean Field Spin Glass Models}, Commun.
Math. Phys. {\bf 230:1}, 71-79 (2002).

\bibitem{gt2} F. Guerra, F. L. Toninelli,
\emph{The high temperature region of the Viana-Bray diluted spin
glass model}, J. Stat. Phys. \textbf{115}, 531-555 (2003).

\bibitem{limterm2} F. Guerra, F. L. Toninelli, {\em
The infinite volume limit in generalized mean field disordered
models}, Markov Processes and Rel. Fields,  {\bf 9}, $195-207$
(2003).

\bibitem{hebb} D.O. Hebb, {\em Organization of Behaviour}, Wiley, New York, 1949.

\bibitem{AI} V. Honavar, L. Uhr (Ed.) {\em Artificial Intelligence and Neural Networks:
Steps Toward Principled Integration}, Elsevier, Boston: Academic
Press (1994).

\bibitem{alb3} J. Hertz, A. Krogh, R. Palmer, {\em Introduction to the theory of neural
computation}, Santa Fe Institute Studies in the Sciences of
Complexity (1991).


\bibitem{hopfield} J.J. Hopfield, {\em Neural networks and physical systems with emergent
collective computational abilities}, Proc. Ntl. Acad. Sci. USA
\textbf{79},  2554-2558 (1982).

\bibitem{moment} M. Krein, A.  Nudelman, {\em The Markov moment problem and extremal problems.
Ideas and problems of P. L. Chebyshev and A. A. Markov and their
further development}, American Mathematical Society, Vol. $50$,
Providence,  1977.

\bibitem{MPV} M. M\'ezard, G. Parisi and M. A. Virasoro, {\em Spin glass theory
and beyond}, World Scientific, Singapore, 1987.


\bibitem{pastur} L. Pastur, M. Shcherbina, {\em The absence of self-averaging
of the order parameter in the Sherrington-Kirkpatrick model}, J.
Stat. Phys. {\bf 62}, 1-19 (1991).

\bibitem{tirozzi2} L. Pastur, M. Scherbina, B. Tirozzi, {\em The
replica symmetric solution of the Hopfield model without replica
trick} J. Stat. Phys. \textbf{74}, 1161-1183 (1994).

\bibitem{tirozzi3} L.Pastur, M. Scherbina, B. Tirozzi, {\em On the replica
 symmetric equations for the Hopfield model} J. Math. Phys. \textbf{40}, 3930-3947 (1999).


\bibitem{talahopfield1}  M. Talagrand, {\em Rigourous results for the Hopfield model with many patterns},
Probab. Th. Relat. Fields \textbf{110}, 177-276 (1998).

\bibitem{talahopfield2} M. Talagrand, {\em Exponential inequalities and convergence of moments in
the replica-symmetric regime of the Hopfield model}, Ann. Probab.
\textbf{38}, 1393-1469 (2000).

\bibitem{Talabook} M. Talagrand, \emph{Spin glasses: a challenge for mathematicians. Cavity and mean field models.},
  Springer Verlag, Berlin, 2003.
  
\bibitem{ALR} Michael Aizenman, Joel Lebowitz, David Ruelle, {\em Some Rigorous Results on the Sherrington-Kirkpatrick Model
of Spin Glasses}, Commun. Math. Phys., \textbf{112} 3-20 (1987).


\end{thebibliography}
\end{document}